# Limits on $CO_2$ Climate Forcing from Recent Temperature Data of Earth

David H. Douglass[a] and John R. Christy[b]

**Abstract**

The global atmospheric temperature anomalies of *Earth* reached a maximum in 1998 which has not been exceeded during the subsequent 10 years. The global anomalies are calculated from the average of climate effects occurring in the tropical and the extratropical latitude bands. *El Niño/La Niña* effects in the tropical band are shown to explain the 1998 maximum while variations in the background of the global anomalies largely come from climate effects in the northern extratropics. These effects do not have the signature associated with $CO_2$ climate forcing. However, the data show a small underlying positive trend that is consistent with $CO_2$ climate forcing with no-feedback.

[a]Department of Physics and Astronomy, University of Rochester, Rochester, NY 14627, USA
[b]Department of Atmospheric Science and Earth System Science Center, University of Alabama in Huntsville, Huntsville, AL 35899, USA

## 1. Introduction

The Intergovernmental Panel on Climate Change (IPCC) has reported that Earth's mean global surface temperature has increased by about *1°C* during the last century [IPCC, 2007]. Estimates of variations in the surface temperature from natural phenomena such as solar, climate shifts from changes in ocean currents, atmospheric aerosols, clouds, changes in albedo, recovery from the "little ice age", conversion of land to agricultural use, *etc.* are comparable in magnitude. A major interest, however, is in the possibility that climate forcing from atmospheric $CO_2$ contributes some part of this increase. The IPCC report also states: "[M]ost of the observed increase in global average temperatures since the mid-20th century is very likely due to the observed increase in anthropogenic greenhouse gas concentrations.". The 'greenhouse gas' is mainly $CO_2$.

Figure 1 shows the global temperature anomalies, *ΔT*, from two commonly used data sets: HadCRUT3 surface measurements and UAH_LT satellite values for the lower troposphere (LT) for the period January 1979 to July 2008. Both data sets show that *ΔT* reached a maximum in 1998 and has not been exceeded in the subsequent 10 years. Also evident are oscillations of period 2 to 5 years which are associated with *El Niño/La Niña* effects (discussed below). The data sets also show differences. The HadCRUT3 values are larger and have a generally increasing background. The MSU values have a smaller trend. This is all discussed below.

### *Climate Theory*

The influence of atmospheric $CO_2$ on the atmosphere is expressed by what is called a climate forcing *ΔF*. The mathematical expression for $CO_2$ is discussed below. In general, climate theory defines ΔF from any source in terms of an equivalent change in net irradiance (in W/m$^2$) referred to the top of the atmosphere [Shine *et al.,* 1995]. This forcing is assumed to causes a change in the mean temperature of the Earth. Climate models define a sensitivity parameter λ relating *ΔF* and *ΔT*

$$\Delta T \approx \lambda \Delta F . \qquad (1)$$

[Note that in some of the earlier literature the symbol for climate sensitivity is the inverse of this IPCC definition.]

The climate sensitivity $\lambda$ can be expressed as the product of two factors

$$\lambda = g\lambda_0, \qquad (2)$$

where $\lambda_0$ is the no-feedback sensitivity and $g$ is the gain resulting from any feedback from the particular climate forcing being considered. For a large number of forcings (including CO2) the no-feedback value is $\lambda_0 = 0.30$ K/(W/m$^2$) [*Kiehl*, 1992; *Shine et al.*, 1995]. There is general agreement among climate scientists for the case of no-feedback. There is disagreement in regard to the validity of the *global warming hypothesis* that states that there are positive feedback processes leading to gains $g$ that are larger than 1, perhaps as large as 3 or 4. However, recent studies suggest that the values of $g$ is much smaller. In a recent study involving aerosols Chylek *et al*. [2007] obtain climate sensitivities of $\lambda = 0.29$ to $0.48$ K/(Wm$^{-2}$). Schwartz (2008) in a study of ocean heat content data reports a smaller value. Also Lindzen *et al*. [1998] and Douglass *et al*. [2005] report low values of $\lambda$ from studies of the Pinatubo volcano event

*Nature of the CO$_2$ response.*

In order to determine if atmospheric $CO_2$ can account for part of the $\Delta T$ variations, it is important to characterize the nature of the $CO_2$ climate forcing. Even though the magnitude of the expected $\Delta T$ signal is yet to be determined, one can, assuming a linear response, make certain assumptions about the signature of the expected $CO_2$ signal.

1. The atmospheric $CO_2$ is slowly increasing with time [Keeling *et al.* (2004)]. The climate forcing according to the IPCC varies as ln ($CO_2$) [IPCC (2001)] (The mathematical expression is given in section 4 below). The $\Delta T$ response would be expected to follow this function. A plot of ln ($CO_2$) is found to be nearly linear in time over the interval 1979-2004. Thus $\Delta T$ from $CO_2$ forcing should be nearly linear in time also.

2. The atmospheric $CO_2$ is well mixed and shows a variation with latitude which is less than 4% from pole to pole [Earth System Research Laboratory. 2008]. Thus one would expect that the latitude variation of $\Delta T$ from $CO_2$ forcing to be also small. It is noted that low variability of trends with latitude is a result in some coupled atmosphere-ocean models. For example, the zonal-mean profiles of atmospheric temperature changes in models subject to "20CEN" forcing ( includes CO2 forcing) over 1979-1999 are discussed in Chap 5 of the U.S. Climate Change Science Program [Karl et al.2006]. The PCM model in Fig 5.7 shows little pole to pole variation in trends below altitudes corresponding to atmospheric pressures of 500hPa.

Thus, changes in $\Delta T$ that are oscillatory, negative or that vary strongly with latitude are inconsistent with $CO_2$ forcing as indicated above.

### *Definition of temperature anomaly*

It is necessary to define temperature $T$ and other quantities describing the climate system of Earth. The radiative-convective equilibrium concept in climate modeling is discussed in a recent National Research Council report [NRC 2005]. In this report, the radiation forcing, the heat content, and the changes in temperature $\Delta T$ are all referenced to the tropopause. Note that the reference is not Earth's surface. Pielke *et al.* [2007] have pointed out that in this context that the $\Delta T$ in the energy balance equations is a "…[t]hermodynamic proxy for the thermodynamic state of the *Earth* system". They then make the point that the surface temperature anomalies are not a good proxy for $\Delta T$ because the measurements are made within the surface boundary layer (SBL) which can in many cases contain effects which result in a decoupling from $\Delta Ts$ higher in the troposphere. Lindzen [2007] makes the same point that the surface temperature anomalies are not a good proxy in a different way. He stresses that the radiation in the energy flux balance relations can be thought of as coming mostly from the atmospheric layer where the infrared optical depth is near 1. This characteristic emission layer (CEL) is above the boundary layer and is typically at an altitude of 7-8km [pressure 400-300hPa] in the tropics.

For these reasons temperature anomalies derived from surface measurements are not a suitable proxy (see also Christy *et al.* 2006). There are additional reasons for not using the surface temperature data that include non-uniform coverage of the globe.

The MSU satellite lower tropospheric (LT) temperature anomalies do cover the globe uniformly and are relatively free from SBL effects because the mean altitude associated with the anomalies is well above that of the SBL. Thus we choose the MSU_LT temperature anomalies as a more suitable proxy. There are, however, two independent groups, University of Alabama in Huntsville (UAH) and Remote Sensing Systems (RSS) which produce different version of LT that are close to each other. The small differences between the two regarding the study at hand do not affect the major conclusions. We choose UAH as the better data set as justified below.

In section 2 we describe the data sets. Section 3 examines the latitude dependence and the causes of the recent variations. Section 4 and 5 give the conclusions and summary.

## 2. Sources of data and methods.

### 2.1. Data sets

All data sets are monthly time series. The time interval of the data is from Jan 1979 to Dec 2007 and is referred to as the satellite era.

#### *Surface temperatures*

The surface temperature measurements are from HadCRUT3. [Jones and Moberg, 2003] This data set is used by the IPCC and by many others. As mentioned above the surface temperature is not a good proxy for the "thermodynamic" temperature that describes the Earth's climate system.

*Microwave sounding units (MSU) data from satellites*

The University of Alabama in Huntsville (UAH) [Christy and Norris, 2006] and Remote Sensing Systems (RSS) (Mears and Wentz, 2005) provide two independent analyses of the same MSU data. The MSU_LT anomalies used in this study represent the lower troposphere (LT) and are a weighted mean from the surface to a pressure of 350 hPa (mean altitude 2.5 km) [Spencer and Christy (1992)]. The importance of the MSU data sets is that all areas of the globe are sampled uniformly. There are small differences between the UAH and RSS data sets which are discussed in appendix A. However, one obtains the same conclusions of this research whichever data set is used. We use the UAH_LT data.

*Latitude bands*. The temperature anomal**y** data can be partitioned into averages over latitude bands that are used in this paper. There are the familiar global (85S-85N) and tropical (20S-20N) latitude bands. North of the equator there are: NH(0-85N), ExTropics (20N-85N), and NoPol (60N-85N). There are corresponding latitude bands south of the equator.

*ENSO* data

Anomalies in the sea surface temperature (*SST*) of particular regions of the *Pacific Ocean* show the *El Niño/ La Niña* phenomena of alternating warm/cold regimes of period 2-5 years. Similar anomalies in pressure are observed across the southern *Pacific* in the *Southern Oscillation Index* (*SOI*). Many investigations have demonstrated correlation between the two phenomena. The general phenomena are called *El Niño Southern Oscillation (ENSO)*. *Barnston et al*. [1997] in a general study with the objective of finding an ENSO index in the tropical Pacific with the strongest correlation with the core *ENSO* phenomena found that a region which they named SST3.4 was best. They introduced a new index, *nino3.4,* straddling the equator [120ºW-170ºW] which *"… [m]ay be regarded as an appropriate general SST index of the ENSO state by researchers, diagnosticians, and forecasters*." The ENSO indices are produced by the Climate Prediction Center of NOAA [NOAA/CPC] Values of the nino3.4 index and others are posted monthly.

*CO₂ data*

We use $CO_2$ concentration values (*C*) measured at *Mauna Loa*. The data are from *Scripps Institution of Oceanography* (*SIO*) [CDAIC, 2007] from 1958 to 2004.

*Aerosol optical density (AOD)*

The AOD index (dimensionless) is generally accepted as the proxy for volcano climate forcing [*Hansen et al.* (2002)]. The most recent determination of AOD is by Ammann *et al.* [2003]. The effects from the two major volcanoes, El Chichón (1982), and Pinatubo (1991) are included in the AOD data tables. The decreases noted by Chylek (2007) are not included in the data tables.

**2.2. Methods and definitions**

In many geophysical data sets an interfering 12-month seasonal effect is a recognized problem. This seasonal effect is "removed" by a variety of schemes before indices of "anomalies" are prepared. However, these methods may not be completely successful. Therefore, we have applied a 12-point trailing average "box" digital filter, *F*, to all time-series considered in this paper. Such a filter is a low pass filter which has the added property of having a zero at a frequency of $1/12$ month$^{-1}$. This filter preserves the monthly resolution of the original time series but will produce a time shift such that all features occur 6 months earlier than the date assigned.

**3. Analysis.**

**3.1 Global**

The global values of *ΔT* in Figure 1 show for the period Jan 1979 to Jan 2008 that the anomalies reached a maximum in 1998 which has not been exceeded by later values. Also evident are the oscillations of 2-5 year period. The global values can be obtained by an average over the three latitude bands: NoExtropics (north of 20N), tropics (20S-20N), and SoExtropics (south of 20S). We show below that climate effects in these latitude bands "stay within the band". To explain the global values one must examine the latitude bands separately. For example, the El Niño/La Niña effects originate in the tropics and are strongest there.

**3.2 Latitude effect**

We have examined the temperature anomalies at the various latitudes enumerated above for three data sets: HadCRUT3v, and MSU_LT from UAH and from RSS. All show similar behavior. However, as explained above, we only present the results from MSU_LT_UAH. Figure 2 shows the UAH_LT anomalies for NoExtropics, Tropics, SoExtropics and Global. The average trends over the range 1979-2007 are 0.28, 0.08, 0.06 and 0.14 ºK/decade respectively. If the climate forcing were only from $CO_2$ one would expect from property #2 a small variation with latitude. However, it is noted that NoExtropics is 2 times that of the global and 4 times that of the Tropics. Thus one concludes that the climate forcing in the NoExtropics includes more than $CO_2$ forcing. These non-$CO_2$ effects include: land use [Peilke *et al.* 2007]; industrialization [McKitrick and Michaels (2007), Kalnay and Cai (2003), DeLaat and Maurellis (2006)]; high natural variability, and daily nocturnal effects [Walters et al. (2007)].

**3.3 The tropical band**

Fig 3 shows the tropical UAH_LT data and the *nino3.4* time-series. One sees that for UAH_LT that the value at the end of the data series is less than at the beginning. However, one should not conclude from this observation that the trend is negative because of the obvious strong correlation between UAH_LT and nino3.4. The exception to this correlation occurs in time-segments following the volcanic eruptions of El Chichón [1982-86] and Pinatubo [1991-95] which cool the troposphere [see Christy and McNider (1994)]. A quick estimate of the magnitude of the correlation can be made by removing the volcano segments and doing a standard correlation calculation. The result

$\qquad$ *UAH*=0.288**nino3.4*+0.044 $\qquad$ (1)

$\qquad$ $R^2$ = 0.864; delay of UAH by 4 months,

where $R^2$ is the coefficient of determination. The correlation of the RSS temperature anomalies vs. *nino3.4* also was studied. The coefficient was nearly the same. However, the value of $R^2$ for *RSS* was 0.678 which is smaller than for UAH. Under the assumption that *ΔT* variations in the tropics are due mainly to *ENSO* then the data set which showed the highest correlation would be best.

This calculation quantifies the high correlation of ΔT and nino3.4 but does not yield the underlying temperature trend. This is determined by multiple regression analysis in the next section.

**3.4 Underlying linear temperature trend**

The expected signature of CO2 climate forcing is a linear time dependence of the temperature anomalies. The global values, however, are not suitable to analyze for that signal because they contains effects from the NoExtropic latitude band which were not consistent with the assumption of how Earth's temperature will respond to $CO_2$ forcing.

Thus we look to the tropical anomalies. If one is able to determine an underlying trend in the tropics, then assuming that the latitude variation of the intrinsic $CO_2$ effect is small ($CO_2$ property #2), then the global trend should be close to this value. The trend, *k*, of the unprocessed tropical data shown in fig 3 is 0.076 *K/decade*. This is obviously not a proper estimate of any underlying trend because of the ENSO effect (nino3.4) and the two volcanoes, El Chichón and Pinatubo, which occurred during this time period

The desired underlying trend, the ENSO effect, the volcano effect can all be determined by a multiple regression analysis [Douglass and Clader (2002)]. The method assumes that *ΔT* depends linearly as

*ΔT = k\*time + $k_1$\*nino3.4+$k_2$\*AOD.* (2)

where the first term is the linear temperature trend, the second is the proxy for ENSO effects and the third term is the proxy for the volcanoes. The trend *k* and the sensitivity coefficients $k_1$, $k_2$ are results which come from the regression analysis. Before beginning the analysis the appropriate time delays must be determined. From the results in section 3, *ΔT* was shown to follow *nino3.4* by 4 months and we determine separately that the delay for *AOD* is 12 months. There is no delay associated with the linear term.

The regression analysis yields $k$, the underlying trend

$$k = 0.062 \pm 0.010 \text{ K/decade.} \qquad (3)$$

The uncertainty is statistical. The coefficient of determination is $R^2 = 0.886$, showing that most of the variance is removed by the regression analysis. The values of the other coefficients from the regression analysis are given in table 1.

There are other systematic climate effects not considered above which could affect the value of the trend, eq3. One example is the solar irradiance which has decreased slightly during this time period. Using results of Douglass and Clader [2002] the effect is estimated to be less than 20%. A second example is from a paper by Chylek *et al.* [2007]. They report a secular decrease in AOD of *-0.0014/year* in recent data. Using the value $k_2 = -2.3$ K that we have found for the AOD sensitivity, one calculates a contribution to the trend of 0.036 K/decade. Although this is a subtraction from the value in eq 3, it is best thought of as one more example of a systematic effect that must be considered. A third effect is black carbon aerosol. Ramanathan and Carmichael [2008] estimate that the climate forcing from absorption of visible light by atmospheric black carbon soot can be as high as 55% of that from $CO_2$. There could be other effects not enumerated. This value, eq 3, is a candidate for a $CO_2$ signal

**4. Discussion and conclusions**

*Warming from $CO_2$ forcing*

How big is the effect from $CO_2$ climate forcing? From IPCC [2001]

$$\begin{aligned}\Delta T(CO_2) &\approx \lambda * \Delta F(CO_2) \\ \Delta F(CO_2) &\approx 5.33 \ln(C/C_0)\end{aligned} \qquad (4ab)$$

where $\lambda$ is the climate sensitivity parameter whose value is 0.30 °K/(Wm$^{-2}$) for no-feedback; $C$ is the concentration of $CO_2$, and $C_0$ is a reference value. From the data the mean value of the slope of $\ln(C(t)/C(t_0))$ vs. time from 1979 to 2004 is 0.044/decade.

Thus,

$$\Delta T(CO_2) \approx 0.070 \text{ °K/decade} \qquad (5)$$

This estimate is for no-feedback. If there is feedback leading to a gain $g$, then multiply eq 5 by $g$.

The underlying trend, eq 3, estimated from the tropical anomalies is consistent with $CO_2$ forcing with no-feedback. It is frequently argued that the gain $g$ is larger than 1, perhaps as large as 3 or 4. This possibility requires there to be some other climate forcing of negative sign to cancel the excess. From the results of Chylek [2007], this cancellation cannot come from aerosols. One candidate is the apparent negative feedback associated with changes in cirrus clouds when warmed [Spencer *et al.* 2007].

### *Is the underlying trend linear?*

Seidel and Lanzante [2004] consider three alternate statistical models for temperature changes different from simple linear trends. Based upon break-points (abrupt changes) the three are: flat steps, piecewise linear and sloped steps. They show that for a number of temperature data sets these models of temperature changes give a better fit. For example, "[R]esults for the tropospheric data suggest that it is reasonable to consider most of the warming during 1958-2001 to have occurred at the time of the abrupt climate regime shift in 1977."

We have considered this possibility for the UAH tropical data in fig 3. Assuming the "flat step" ('flat' means *0* slope) model with just one step we find a unique solution. There is a step in 1997 of magnitude of $\approx 0.2$ K. The choice between a model of a linear trend and one with abrupt changes depends on ones understanding of the measurement techniques and physical processes of the climate system. Randal and Herman [2008], in fact, used such a breakpoint analysis of measurement techniques to conclude that the UAH temperature data is better than that of RSS. In the appendix, we find one such break-point in the RSS temperature data.

## 5. Summary

The recent atmospheric global temperature anomalies of the *Earth* have been shown to consist of independent effects in different latitude bands. The tropical latitude band variations are strongly correlated with ENSO effects. The maximum seen in 1998 is due to the *El Niño* of that year. The effects in the northern extratropics are not consistent with $CO_2$ forcing alone

An underlying temperature trend of $0.062 \pm 0.010 °K/decade$ was estimated from data in the tropical latitude band. Corrections to this trend value from solar and aerosols climate forcings are estimated to be a fraction of this value. The trend expected from $CO_2$ climate forcing is *0.070g °C/decade*, where *g* is the gain due to any feedback. If the underlying trend is due to $CO_2$ then *g*~1. Models giving values of *g* greater than *1* would need a negative climate forcing to partially cancel that from $CO_2$. This negative forcing cannot be from aerosols.

These conclusions are contrary to the IPCC [2007] statement: "[M]ost of the observed increase in global average temperatures since the mid-20th century is very likely due to the observed increase in anthropogenic greenhouse gas concentrations."

**Acknowledgements.**

We wish to thank R. S. Knox and J.J. Hnilo for helpful discussions.

**Appendix A. Comparison of MSU and RSS**

The University of Alabama in Huntsville (UAH) [Christy and Norris, 2006] and Remote Sensing Systems (RSS) (Mears and Wentz, 2005) provide two independent analyses of the same MSU data[1979-2007]. The MSU_LT anomalies used in this study represent the lower troposphere (LT) and are a weighted mean from the surface to a pressure of 350 hPa (mean altitude 2.5 km) [Spencer and Christy (1992)]. The importance of the MSU data sets is that all areas of the globe are sampled uniformly. A weakness is that the record does not begin until 1979.

Randall and Herman [2008] report a detailed comparison of UAH and RSS in an effort to determine the causes of the discrepancies between the two data sets. They found that the discrepancies were associated with adjustments from one satellite to another and with diurnal corrections. Comparison with radiosonde data sets "… [i]ndicated that RSS's method … of determining diurnal effects is likely overestimating the correction to the LT channel." In other words, Randall and Herman state that the RSS methods lead to warm biases and thus the UAH data set is likely better. In particular, they state that the largest discrepancies [RSS larger than UAH] in the LT channel are centered on 1993 in both the global and tropical data. There are also other smaller discrepancies.

Christy and Norris [2006] and Christy *et al.* [2007] provide additional evidence that UAH is preferred over RSS. Their conclusions are based upon (a) An examination of specific time periods (b) A study of the inter-relationships between MSU bulk layer temperatures and (c) In a comparison with a uniform dataset of U.S. radiosondes, RSS tropospheric temperatures revealed a significant upward shift of about 0.1 K in the early 1990s. Further comparisons with tropical radiosondes and surface temperature datasets indicated the same result, that in comparison with all others, RSS displayed a relative positive shift of 0.07 to 0.13 K. In absolute terms, RSS was the only tropical tropospheric dataset which indicated 3-year average temperatures were significantly warmer after the eruption of Mt. Pinatubo than before. Finally, in a test of inter-layer consistency (i.e. the relationship between temperatures of satellite products measuring different vertical layers), RSS data were outside the statistical estimates calculated by radiosonde measurements (Christy et al. 2007).

In the text of this paper we showed that the anomalies in the tropics are strongly correlated with ENSO and since ENSO effects obviously have no break-points or diurnal corrections, then the data set that best processed the break points and diurnal corrections would have the highest correlation with nino3.4. UAH had the larger $R^2$.

Can we determine where the differences between UAH and RSS are? And their magnitude? Since RSS has the more positive linear trend, published evidence shows that there is a "jump" between the two data sets sometime during the early-mid 1990s. This possibility was tested on the tropical data. In particular, the total time-segment was divided into an early period and a late period separated by a short time-segment that was removed. Fig A1 shows a plot of RSS vs. UAH. Blue is the early time-segment and red is the late time-segment. The beginning and end of the removed segment were varied to give the largest coefficient of determination, $R^2$, while keeping the slope near *1*.This procedure leads to a unique removed-segment from mid-92 to mid-94 (see Christy and Norris [2006], Christy *et al.* [2007] and Randal and Hermann [2008] for more detail). The jump was 0.136ºK. This and other results are tabulated in table S1.

By these tests we view UAH as the better data set

**Figure Captions**

Fig 1.   Global temperature anomalies for period 1979-2007 for the satellite UAH_LT and the surface HadCRUT3.

Fig.2.  UAH_LT temperature anomalies: northern hemisphere, southern hemisphere, tropics and global from 1979-2007.

Fig. 3.  UAH_LT  tropical temperature anomalies and ENSO index, nino3.4, from 1979 to 2007.

Fig A1. Comparison of tropical  UAH_LT  and RSS_LT data sets from 1979 to 2007

| Table 1. Multiple regression analysis of UAH tropical $\Delta T$ anomalies. $\Delta T = k*time + k_1*nino3.4 + k_2*AOD$. For the values below, the coefficient of determination is $R^2 = 0.886$ ||||
|---|---|---|---|
| predictor | linear time | nino3.4 | AOD |
| symbol | $k$ | $k_1$ | $k_2$ |
| units | ºK/year | ºK/ºK | ºK/unit AOD |
| value of coefficient | 0.00620±0.0010 | 0.281±0.012 | -2.60±0.24 |
| delay (months) | na | 4 | 12 |

| Table A1. Comparison of tropical UAH and RSS ||||
|---|---|---|---|
|  | Early period | Late period |  |
| Time segment | 1979 to mid-1992 | Mid-1994 to 2007 | 2 year segment removed |
| equation | *RSS=0.998*UAH-0.023* | *RSS=1.00*UAH+0.113* | Jump of 0.136ºK |
| $R^2$ | 0.985 | 0.975 |  |

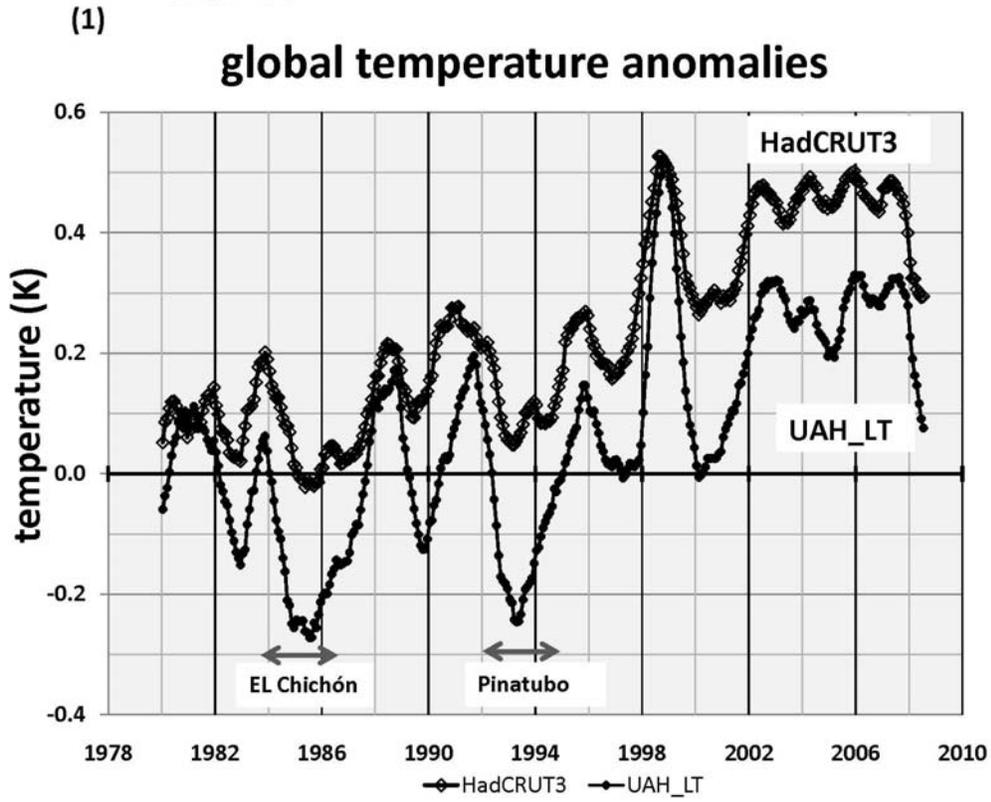

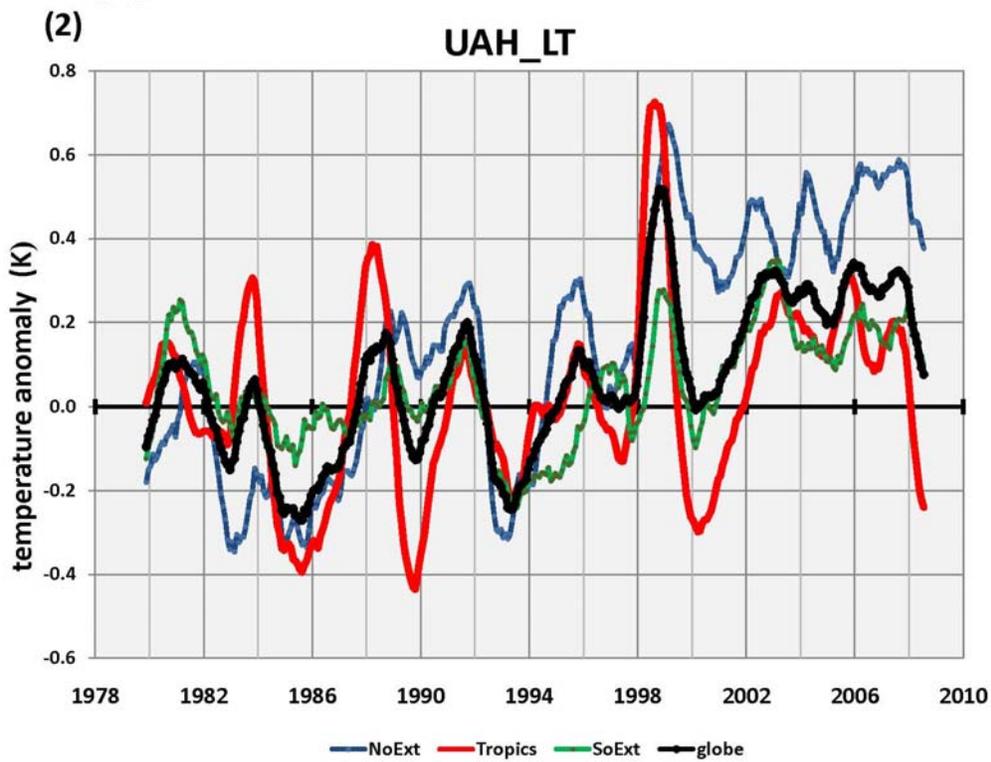

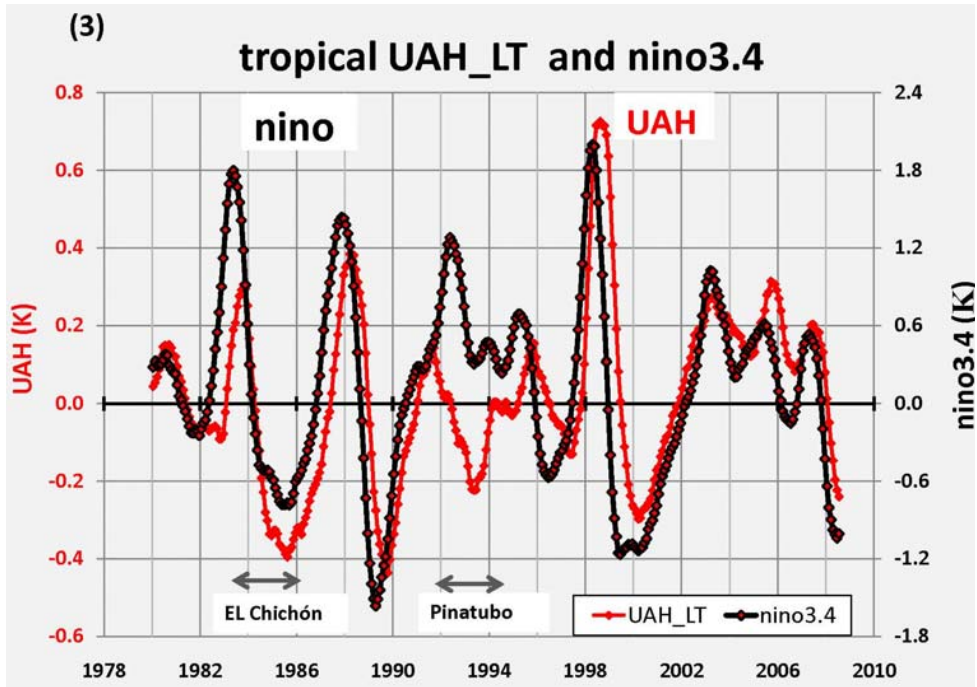

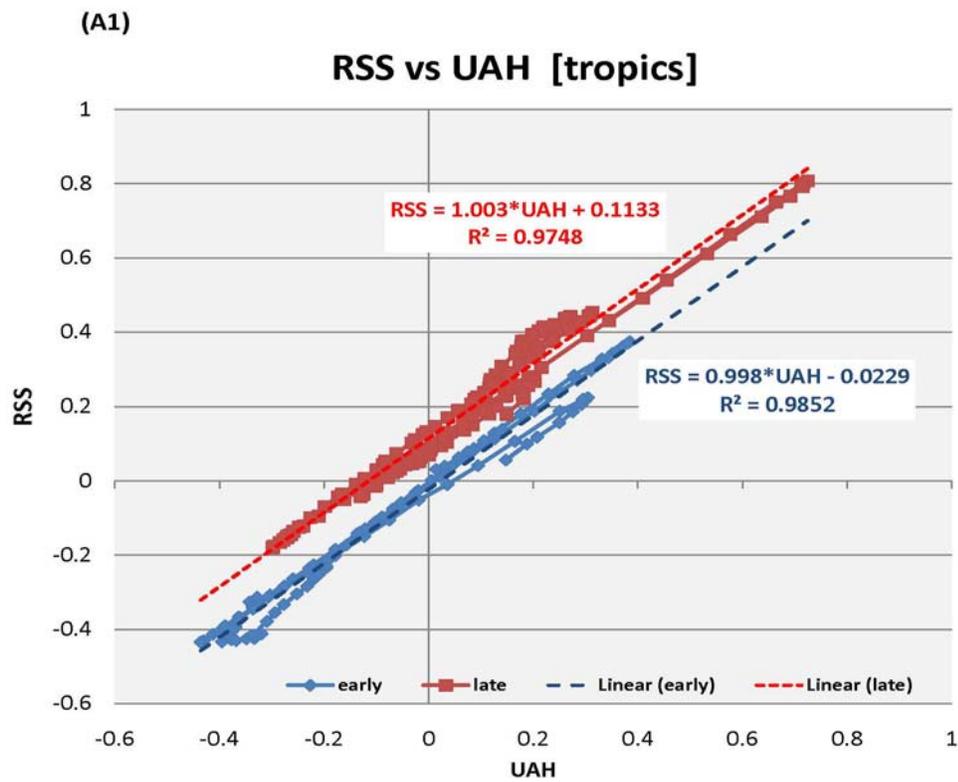